\documentstyle[aps,prl,multicol,epsfig]{revtex}

\renewcommand{\narrowtext}{\begin{multicols}{2} \global\columnwidth20.5pc}
\renewcommand{\widetext}{\end{multicols} \global\columnwidth42.5pc}
\renewcommand{\v}[1]{{\bf #1}}

\newcommand{\eqa}{\begin{eqnarray}}
\newcommand{\eea}{\end{eqnarray}}
\newcommand{\eq}{\begin{equation}}
\newcommand{\ee}{\end{equation}}

\begin{document}
\draft
\title{The Superfluid Density in the D-density-wave Scenario}
\author{Qiang-Hua Wang$^{1,2}$, Jung Hoon Han$^1$, and Dung-Hai Lee$^1$}
\address{$^1$Department of Physics,University of California at Berkeley,
Berkeley, CA 94720, USA}
\address{$^2$Physics Department and National Laboratory of Solid State Microstructures, \\
Institute for Solid State Physics, Nanjing University, Nanjing 210093, China}
\maketitle

\begin{abstract}
\rightskip 54.8pt Recently Chakravarty, Laughlin, Morr and Nayak made an
interesting proposal that the cuprate superconductors possess a hidden
``d-density wave'' order. We study the implication of this proposal for the
superfluid density $\rho_s$. We find that it predicts a temperature gradient 
$|d\rho_s/dT|_{T=0}$ that is strongly doping dependent near the critical
doping at which the superconducting gap vanishes.
\end{abstract}


\narrowtext

Recently Tallon and Loram critically examined the existing experimental data
on specific heat, photoemission, magnetic susceptibility and optical
conductivity of the cuprate superconductors\cite{tl}. They argued that these
data are consistent with the existence of two competing energy gaps, a
pseudogap and the superconducting gap, both of which disperse according to
d-wave symmetry. The pseudogap exists in the normal state of the underdoped
cuprates while the superconducting gap opens up in the BCS fashion at
temperature $T_c$. Tallon and Loram also argue that the data indicates the
closing of the pseudogap at a doping level $x\approx 19\%$.

This observation motivated Chakravarty, Laughlin, Morr and Nayak (CLMN) to
propose the ``d-density-wave'' (DDW) scenario\cite{clmn}.
According to this proposal a staggered long-range order in the orbital
magnetic moment, as firstly suggested by Hsu, Marston and Affleck in another
context,\cite{affleck}, exists at low temperatures when the doping is less than a critical value $
x_c\approx 0.19$. Such order gives rise to a non-superconducting gap (the
pseudogap) with the same $(\cos{k_x}-\cos{k_y})$ dispersion as the
superconducting (SC) gap. Moreover, the DDW order competes with
superconductivity and causes $T_c(x)$ to trace out the familiar
``superconducting dome'' in the doping-temperature phase diagram of the
cuprates. So far there is no direct experimental evidence for the DDW order.
Neither is there evidence for the expected Ising-like phase transition into
such a state. CLMN argue that disorder gets in the way of a sharp phase
transition and turns it into a cross-over.

In addition to offering a possible new phase for high-$T_c$ cuprates, there
are a number of attractive features in CLMN's proposal. Since the DDW metal
(i.e. the normal state in the presence of DDW order) is a doped
band-insulator one expects the Drude weight in the normal state\cite{drude},
and the superfluid density in the superconducting state to be both
proportional to the doping density\cite{wl}, as indeed found in the
experiments. The closing of the DDW gap at $x=0.19$ can easily explain the
observed kink in the jump of the T-linear specific heat coefficient
\cite{loram}. In addition the small hole pockets centered around the nodes of the
DDW gap could be the progenitor of the Fermi arcs observed in angle-resolved
photoemission\cite{ding}. From a technical point of view a doped band
insulator has more resemblance to a doped Mott insulator than a
large-Fermi-surface metal, hence is a better starting point for description
of the underdoped cuprates. Finally the CLMN proposal is crisp and in
principle falsifiable. For all the above reasons we feel that it is 
worthwhile to further check the prediction of this proposal against existing
experiments.

The theory presented in Ref.\cite{clmn} is mean-field in nature. For such a
theory, quasiparticles are sharply defined. Since a much better case can be
made for the existence of quasiparticles in the superconducting states
\cite{arpes}, we confine our calculations to low temperatures where SC order
parameter exist.

Among various superconducting properties we focus on the superfluid
density  $\rho _{s}$ because of its rather unconventional doping and
temperature dependence, which has been the focus of many theories
\cite{wl,levin}. Experimentally it is established that, for a fairly wide
range of doping, $d\rho _{s}/dT$ is nearly doping independent at low
temperatures\cite{bonn,lemberger}. In contrast the extrapolated zero
temperature superfluid density changes significantly with doping\cite
{lemberger}. The goal of this paper is to work out the DDW theory's
prediction for $\rho _{s}(T,x)$ (in the absence of disorder).

Even if the quasiparticles are well-defined in the underdoped
superconducting states, the Mott constraint can substantially modify the
results of the free, mean-field theory. Therefore along the mean-field
prediction we also present the results of the ``projected DDW model'' where
the electron occupation constraint is taken into account. More specifically,
the no-double-occupancy constraint is implemented by introducing the slave
bosons (holons) plus the gauge fields which couple to both bosons and
fermions (spinons). The strict occupation constraint is reflected in the
absence of the Maxwell term for the gauge fields, i.e. the coupling constant
is infinity. In a recent paper, one of us looked into such a gauge theory
where the underlying mean-field vacuum is the $d$-wave RVB state of Kotliar
and Liu\cite{kl}. It was shown that the gauge field can be integrated out
exactly in the continuum approximation of the lattice theory\cite{dhl,nayak}%
. This continuum theory describes the (correlated) density and current
fluctuations of holons and spinons above the length scale of the inter-holon
distance $\lambda_h$. The physics below such a length scale is entirely
summarized by a few parameters in the effective action. In the absence of
more accurate estimates for these parameters, the author of Ref.\cite{dhl}
took the mean-field prediction of them.

In the following treatment of the projected DDW model we shall follow the
same path taken in Ref.\cite{dhl} while replacing the Kotliar-Liu RVB
mean-field vacuum by the CLMN mean-field vacuum in the spinon sector. We
expect the program carried out in Ref.\cite{dhl} to work well in the
presence of DDW order, because the DDW metal-a doped band insulator-has very
little spinon density and current fluctuations below the length scale $%
\lambda _{h}$. The fluctuatioins above this length scale is already captured
by the analysis of Ref. \cite{dhl}.

Our results are as follows. For $x_l \!<\! x \!<\! x_u$ and at low
temperatures, the mean-field DDW theory predicts a superfluid density that
behaves as 
\begin{equation}
\rho _{s}(T,x)=\rho _{ddw}(0,x)-\alpha _{ddw}(x)T.  \label{ddwr}
\end{equation}
Results for $\rho _{ddw}(0,x)$ are shown in the main panel, Fig. 1(a). Note
that the zero-temperature superfluid density is non-zero at $x=x_l, x_u$. An
infinitesimal temperature will, however, destroy superfluidity because the
pairing gap vanishes there. The temperature gradient of $\rho _{s}$ is given
by $\alpha_{ddw}(x)\sim t/\Delta (x)$ where $t$ is the hopping integral and $%
\Delta (x)$ is the maximum {\it d}-wave superconducting gap. It diverges at
two points, $x=x_l$, and $x=x_u$ because $\Delta(x)$ vanishes there. In
general the doping dependence of $\Delta (x)$ is reflected in $\alpha
_{ddw}(x)$ as shown in the insert of Fig. 1(a).

The projected DDW model predicts 
\begin{equation}
\rho _{s}(T,x)=z_j(x)\rho _{ddw}(0,x)-z_{j}(x)^{2}\alpha_{ddw}(x)T,
\label{projr}
\end{equation}
where $z_{j}(x)$ is the current renormalization factor to be explained
later. The doping dependence of $z_j(x)\rho _{ddw}(0,x)$ (main panel) and $%
z_{j}(x)^{2}\alpha _{dds}(x)$ (insert) are shown in Fig.1(b). Unlike the
mean-field result, $d\rho _{s}/dT$ has a sharp variation near $x_c$ which
originates from the kink in the mean-field superfluid density $%
\rho_{ddw}(x,0)$.

Let us compare the above theoretical results with existing data on $%
d\rho_{s}(T,x)/dT$. In the oxygen-depleted YBCO thin films reported in Ref.
\cite{lemberger}, the transition temperature $T_{c}$ ranges from $90K$ at
the optimal doping to $38K$ on the underdoped side. Meanwhile, $|d\rho
_{s}/dT|$ decreases by about $15\%$ from its {\it maximal} value at optimal
doping. In contrast, the mean-field DDW theory predicts an increase of $%
|d\rho _{s}/dT|$ by $140\%$, provided $T_{c}$ scales with $\Delta _{0}(x)$.
After projection, the dependence of $|d\rho _{s}/dT|$ on $x$ is about 
$100\%$
.
It should be pointed out that in the DDW theory $|d\rho_{s}/dT|$ is not
universal, and is subject to quantitative change if the doping and
temperature dependences of the pairing amplitude are different from what we
assumed in this paper. It is not out of question that the experimental data
happens to span the plateau regime in Fig.1(b) (insert). In this case, the
present theory predicts a rapid increase of $|d\rho_{s}/dT|$ when the doping
level is decreased even further. In the following we report the details of
the calculations. 
\begin{figure}
\epsfxsize=9cm \epsfysize=8cm \vskip -4cm \epsfbox{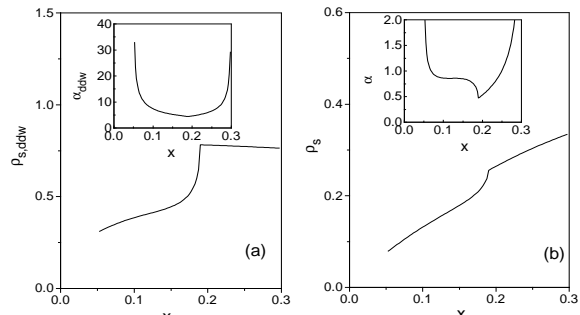}
\caption{(a) The doping dependence of $\protect\rho_{s,ddw}(0,x)$ (main
pannel) and $\protect\alpha_{ddw}(x)$ (inset). (b) The doping dependence of $\protect\rho_{s}(0,x)=z_j (x)\protect\rho_{ddw}(0,x)$ and $\protect\alpha
(x)=z_{j}(x)^{2}\protect\alpha_{ddw}(x)$ using the same notation as in (a).
The parameters used are $\Delta_{0}=0.2t$, $D_{0}=t$, and $t_{b}=2t$.}
\end{figure}

\noindent {\bf The mean-field DDW theory} Following CLMN, we adopt the
following mean-field Hamiltonian 
\begin{eqnarray}
H_{DDW} &=&\sum_{k\sigma }(X_{k}-\mu )c_{k\sigma }^{\dagger }c_{k\sigma
}+\sum_{k\sigma }(iD_{k}c_{k+Q\sigma }^{\dagger }c_{k\sigma }+{\rm h.c.}) 
\nonumber \\
&-&\sum_{k}\Delta _{k}(c_{k\uparrow }c_{-k\downarrow }+{\rm h.c.}).
\label{mfh}
\end{eqnarray}
Here $Q=(\pi ,\pi )$, and $X_{k}=-2t(\cos k_{x}+\cos k_{y})$ is the nearest-
neighbor tight-binding dispersion relation, $D_{k}=D(x)(\cos k_{x}-\cos
k_{y})$ is the momentum space DDW order parameter, and $\Delta _{k}=\Delta
(x)(\cos k_{x}-\cos k_{y})$ is the $d$-wave superconducting gap function. In
the above and the rest of the paper we shall assume that the lattice
constant is unity.

Due to the breaking of translation symmetry by the DDW order, the Brillouin
zone is half of its original size. For $\Delta (x)=0$ there are two bands
given by $\varepsilon _{k\pm }=\pm \sqrt{X_{k}^{2}+D_{k}^{2}}-\mu $\cite{comment}.
In the  presence of pairing, the quasi-particle dispersion becomes 
$E_{k\pm }=
\sqrt{\varepsilon _{k\pm}^{2}+|\Delta _{k}|^{2}}$.

In order to study the doping dependence of the superfluid density it is
necessary to specify the dependence of $D(x)$ and $\Delta(x)$ on $x$.
Following CLMN, we use the solution of $\partial E(D,\Delta)/\partial D=0$
and $\partial E(D,\Delta)/\partial\Delta=0$ to parametrize $D(x)$ and $%
\Delta(x)$ where 
\begin{eqnarray}
E(D,\Delta)&&=a_D(x-x_d)D^2+a_{\Delta}(x-x_u)\Delta^2+b_{D}D^{4}  \nonumber
\\
&&+b_{\Delta }\Delta^4+wD^2\Delta^2.  \label{ef}
\end{eqnarray}
Here $a_{D}$, $a_{\Delta}$, $b_D$, $b_{\Delta}$, and $w$ are (positive)
material-dependent constants, while $x_d$ ($x_u$) is the doping level below
which DDW (SC) order parameter become nonzero in the absence of the other.
In the presence of pairing the onset of DDW happens at $x_c < x_d$ while the
SC order parameter begins its decline precisely at $x_c$ and vainshes at $x_l
$. The functional form for the order parameters are given by 
\begin{eqnarray}
D(x)&=&D_{0}\sqrt{(x_{c}-x)/x_{c}}  \nonumber \\
\Delta(x)&=&\Delta_{0}\sqrt{(x-x_{l})/(x_{c}-x_{l})}; \,\,\,x_l<x<x_c 
\nonumber \\
\Delta(x)&=&\Delta _{0} \sqrt{(x_{u}-x)/(x_{u}-x_{c})}; \,\,\, x_c<x<x_u.
\end{eqnarray}
Here $D_0$ ($\Delta_0$) is the overall scale of $D$ ($\Delta$). In the
following we set $x_l=0.05$, $x_{c}=0.19$, $x_d=0.2$ and $x_u=0.3$.

\widetext
The superfluid density meausures the free energy increase caused by the phase
twist of the superconducting order parameter\cite{feynman}.
After some lengthy but straightforward algebra we obtain
\begin{eqnarray}
\rho _{s}^{ab} &=&{\frac{1}{L^{2}}}\sum_{k,\nu =\pm }\left( 1-\frac{%
\varepsilon _{k\nu }}{E_{k\nu }}\tanh \frac{\beta E_{k\nu }}{2}\right)
\partial _{a}\partial _{b}\varepsilon _{k\nu }-{\frac{\beta }{2L^{2}}}%
\sum_{k,\nu =\pm }\partial _{a}\varepsilon _{k\nu }\partial _{b}\varepsilon
_{k\nu }{\rm sech}^{2}\frac{\beta E_{k\nu }}{2}  \nonumber \\
&&+{\frac{4}{L^{2}}}\sum_{k,\nu =\pm }\frac{\nu |\Delta _{k}|^{2}}{%
E_{k+}^{2}\!-\!E_{k-}^{2}}\frac{\tanh (\beta E_{k\nu }/2)}{E_{k\nu }}\frac{%
(X_{k}\partial _{a}D_{k}\!-\!D_{k}\partial _{a}X_{k})(X_{k}\partial
_{b}D_{k}\!-\!D_{k}\partial _{b}X_{k})}{X_{k}^{2}\!+\!D_{k}^{2}},
\label{form}
\end{eqnarray}
\narrowtext 
\noindent 
where $\partial _{a}$ denotes the momentum derivative in the 
$a$-direction, $L$ is the linear dimension of the system. By symmetry, the
tensor is diagonal and direction-independent. The last term is non-zero only
if the DDW and superconducting order coexist.
It is a consequence of the time-reveral symmetry breaking.
A similar expression, without the last term, has been derived in
Ref. \cite{comment}.  For the entire doping range
considered, the third term has a negligible contribution to the superfluid
density compared to others. For self-consistency, one can check that the
superfluid density vanishes in the absence of pairing due to the fact that
1) the third term vanishes identically when $\Delta (x)=0$. 2) The first two
terms combine to give $\sum_{k\nu }\partial _{a}\{[1\!-\!\tanh (\beta
\varepsilon _{k\nu }/2)]\partial _{b}\varepsilon _{k\nu }\}=0$. At
temperatures much smaller than the maximum superconducting gap, the
suppression of superfluid density comes from the thermal excitation of nodal
quasiparticles. The second term in Eq.~(\ref{form}) gives a $T$-linear
suppression, while all the other terms contribute higher order temperature
corrections.\ As a result, the superfluid density takes the form of Eq.~(\ref
{ddwr}).

We find $\alpha_{ddw}(x)=(4\ln {2})t/\pi \Delta (x)$, independent of the DDW
order parameter. As to $\rho _{ddw}(0,x)$ in Eq.~(\ref{ddwr}), we were only
able to compute it numerically, and the result is shown in Fig.1(a).

The general trend of the doping dependence of $\rho_{ddw}(0,x)$ shown in
Fig.1(a) is consistent with experimental data. However the same cannot be
said about $\alpha_{ddw}(x)$. Due to the competition between DDW order and
superconductivity in the DDW theory, $\Delta(x)$ is suppressed by the
emergence of the DDW order, forming a dome. Thus $\Delta(x)\rightarrow 0$ as 
$x\rightarrow x_l,x_u$ which implies that $\alpha_{ddw}(x)$ diverges as $%
x\rightarrow x_{l}, x_u$. Such $x$-dependent $d\rho_s/dT$ has not been
observed experimentally. \widetext
\noindent {\bf The projected DDW model} The lattice action for the projected
DDW model is given by 
\begin{eqnarray}
L &=&L_{b}+L_{DDW}-i\sum_{i}a_{0i}  \nonumber \\
L_{b} &=&\sum_{i}[\bar{b}_{i}(\partial _{0}\!+\!ia_{i0}\!-\!iA_{i0}\!-\!\mu
_{b})b_{i}]-t_{b}\sum_{\langle ij\rangle }[{\rm e}^{i(a_{ij}-A_{ij})}\bar{b}%
_{i}b_{j}+{\rm h.c.}]+U_{b}\sum_{\langle ij\rangle }\bar{b}_{i}b_{i}\bar{b}%
_{j}b_{j}  \nonumber \\
L_{DDW} &=&\sum_{i}[\bar{f}_{i{\sigma }}(\partial _{0}\!+\!ia_{i0}\!-\!\mu
_{f})f_{i{\sigma }}]-\sum_{\langle ij\rangle }[(t\!+\!iD_{ij}){\rm e}%
^{ia_{ij}}\bar{f}_{i{\sigma }}f_{j{\sigma }}\!+\!{\rm h.c.}]-\sum_{\langle
ij\rangle }[\Delta _{ij}{\rm e}^{i\phi _{ij}}\epsilon _{{\sigma }{\sigma }%
^{\prime }}\bar{f}_{i{\sigma }}\bar{f}_{j{\sigma }^{\prime }}+{\rm h.c.}]
\label{pddw}
\end{eqnarray}
\narrowtext 
\noindent In the above $b_{i}$ and $f_{i{\sigma }}$ are the holon and spinon
fields respectively. With $\phi _{ij}$ set to zero, $L_{DDW} $ is the real
space equivalent of Eq.~(\ref{mfh}). In Eq.~(\ref{pddw}) $a_{\mu }$ is the
gauge field that enforces the constraint $b_{i}^{\dag }b_{i}+f_{i{\sigma }%
}^{\dag }f_{i{\sigma }}=1.$ The DDW mean-field theory describes a doped band
insulator, consequently spinon density fluctuations only occurs above the
length scale $\lambda _{h}$. Obviously the same is true for the holon
density fluctuation. As a result it should be adequate to project out the
spinon and holon density fluctuations above the length scale $\lambda _{h}$.

The effective action above such cutoff length is 
\begin{eqnarray}
&&{\cal L}={\cal L}_{b}+{\cal L}_{fp}+{\cal L}_{Dirac}+{\cal L}_{j} 
\nonumber \\
&&{\cal L}_{b}=\frac{K_{b}}{2}|\phi _{b}^{\ast }({\nabla }+i\v {a}-i \v {A}%
)\phi _{b}|^{2}+\frac{U_{b}}{2}\delta \rho _{b}^{2}  \nonumber \\
&&{\cal L}_{sp}=\frac{K_{sp}}{2}|\phi _{sp}^{\ast }({\nabla }+2i\v {a} )\phi
_{sp}|^{2}+\frac{1}{2U_{sp}}(\phi _{sp}^{\ast }\partial _{0}\phi
_{sp}+2ia_{0})^{2}  \nonumber \\
&&{\cal L}_{j}=i\delta \rho _{b}(\phi _{b}^{\ast }\partial _{0}\phi
_{b}+ia_{0}-iA_{0})+J_{\mu }^{qp}(\phi _{sp}^{\ast }\partial _{\mu }\phi
_{sp}+2ia_{\mu })  \nonumber \\
&&~~~~~~+{\bar{\rho}}(\phi _{b}^{\ast }\partial _{0}\phi _{b}-\frac{1}{2}%
\phi _{sp}^{\ast }\partial _{0}\phi _{sp})-i{\bar{\rho}}A_{0}-i\v {j}%
_{0}\cdot \v {A}.  \label{hyg}
\end{eqnarray}
Here ${\bar{\rho}}$ is the doping density, and $\v {j}_{0}$ is the
transverse ground-state current produced by the DDW order, $\phi _{b}$ and $%
\phi _{sp}$ are the U(1) phase factors associated with the holon field and
the spinon pair-field respectively. In addition ${\cal L}_{Dirac}$ is the
Dirac action for the spinon quasiparticles near the d-wave gap nodes, $%
J^{qp}=\frac{1}{2}(\sum_{n}\psi _{n{\sigma }}^{\dagger }\tau _{z}\psi _{n{%
\sigma }},iv_{F}\psi _{1{\sigma }}^{\dagger }\psi _{1{\sigma }},iv_{F}\psi
_{2{\sigma }}^{\dagger }\psi _{2{\sigma }})$ is their 3-current ($\tau _{z}$
is the third component of the Pauli matrices, and $\psi _{n{\sigma }}$ is
the spinon Nambu spinor associated with the nth d-wave gap node). $%
K_{b}=t_{b}x$ is the holon superfluid density, $4K_{sp}$ is the spinon
zero-temperature superfluid density given by Eq.~(\ref{ddwr}). $U_{b}$ and $%
U_{sp}$ also depend on the parameters in Eq.~(\ref{pddw}), however their
values are not important for the following discussion.

Given Eq.~(\ref{hyg}) the gauge field $a_{\mu }$ can be integrated out
straightforwardly to yield the final effective action of a correlated DDW
superconductor 
\begin{eqnarray}
{\cal L} &=&\frac{K}{2}|\phi ^{\ast }{\nabla }\phi -2i\v {A}|^{2}+\frac{1}{2U%
}(\phi ^{\ast }\partial _{0}\phi -2iA_{0})^{2}  \nonumber \\
&&-z_{j}\v {J}^{qp}\cdot (\phi ^{\ast }{\nabla }\phi -2i\v {A})-z_{\rho
}\rho ^{qp}(\phi ^{\ast }\partial _{0}\phi -2iA_{0})  \nonumber \\
&&+\frac{\bar{\rho}}{2}\phi ^{\ast }\partial _{0}\phi -i{\bar{\rho}}A_{0}-i%
\v {j}_{0}\cdot \v {A}+{\cal L}_{Dirac}^{\prime }.  \label{main}
\end{eqnarray}
In the above $\phi \equiv \phi _{sp}^{\ast }\phi _{b}^{2}$, and 
\begin{eqnarray}
&&K\equiv K_{sp}K_{b}/(K_{b}+4K_{sp}),~~~U\equiv U_{sp}+4U_{b}  \nonumber \\
&&z_{j}\equiv K_{b}/(K_{b}+4K_{sp}),~~~z_{\rho }\equiv
U_{sp}/(U_{sp}+4U_{b}),  \label{ren}
\end{eqnarray}
and 
\begin{equation}
{\cal L}_{Dirac}^{\prime }\equiv {\cal L}_{Dirac}+\frac{2|\v {J}^{qp}|^{2}}{%
K_{b}+4K_{sp}}+\frac{2(\rho ^{qp})^{2}}{U_{b}^{-1}+4U_{sp}^{-1}}.
\label{qp1}
\end{equation}
Combining the above result with Eq.~(\ref{ddwr}) we obtain Eq.~(\ref{projr})
as the superfluid density prediction of the projected DDW theory where 
\begin{equation}
z_{j}(x)=t_{b}x/[t_{b}x+\rho _{ddw}(0,x)].
\end{equation}
In Fig 1(b) we plot the $x$-dependence of the zero-temperature superfluid
density in the main pannel and $d\rho _{s}/dT$ in the insert. For $x>x_{l}$
the zero-temperature superfluid density varies with $x$ in roughly linear
fashion. As in the mean-field DDW theory, $d\rho _{s}/dT$ has a strong $x$%
-dependence.

In conclusion, we showed that in the DDW scenario the zero-temperature
superfluid density increases monotonically with doping. Its temperature
gradient $|d\rho_s /dT|$, being proportional to $1/\Delta(x)$, also depends
sensitively on doping. Such sensitivity to doping is expected in any
mean-field theory where another order parameter competes with $d$-wave
superconductivity. It is possible, however, that this result is due to the
particular approximation scheme employed in this paper, and that another
scheme may well render the temperature gradient insensitive to doping.

\acknowledgments{We thank Sudip Chakravarty for a critical reading and comments
of the manuscript. This work was initiated
during the high-$T_c$ workshop at the Institute for Theoretical Physics, UCSB. DHL is supported by NSF grant DMR
99-71503. QHW is supported by the National Natural Science Foundation of China and the Ministry of Science and
Technology of China (NKBSF-G19990646), and in part by the Berkeley Scholars Program.}

\widetext


\begin{references}
\bibitem{tl}  J. L. Tallon and J. W. Loram, cond-mat/0005063.

\bibitem{clmn}  S. Chakravarty, R. B. Laughlin, D. K. Morr and C. Nayak,
cond-mat/0005443.

\bibitem{affleck} T. C. Hsu, J. B. Marston, and I. Affleck, Phys. Rev. B
{\bf 43}, 2866 (1991).

\bibitem{drude}  J. Orenstein {\it et al}, Phys. Rev. B {\bf 42}, 6342
(1990).

\bibitem{wl}  P. A. Lee and X.-G. Wen, Phys. Rev. Lett. {\bf 78}, 4111
(1997); P. A. Lee {\it et al.}, Phys. Rev. B {\bf 57}, 6003 (1998).


\bibitem{loram}  J. L. Tallon, {\it et al}, Phys. Stat. Sol. B {\bf 215},
531 (1999); J. W. Loram, {\it et al}, J. Phys. Chem. Solids {\bf 59}, 2091
(1998).

\bibitem{ding}  A. G. Loeser, {\it et al}, Science {\bf 273}, 325 (1996); D.
S. Marshall {\it et al}, Phys. Rev. Lett. {\bf 76}, 4841 (1996); J. M.
Harris {\it et al}, Phys. Rev. Lett. {\bf 79}, 143 (1997).

\bibitem{arpes}  A. G. Loeser {\it et al}, Phys. Rev. B {\bf 56}, 14185
(1997); A. V. Fedorov {\it et al}, Phys. Rev. Lett. {\bf 82}, 217 (1999); A.
Kaminski {\it et al}, Phys. Rev. Lett. {\bf 84}, 1788 (2000).

\bibitem{levin} Q. Chen {\it et al.}, Phys. Rev. Lett. {\bf 81}, 4708 (1998).

\bibitem{bonn}  D. A. Bonn {\it et al}, Czech J. Phys. {\bf 46}, S6. 3195
(1996).

\bibitem{lemberger}  B. R. Boyce, K. M. Paget and T. R. Lemberger,
cond-mat/9907196, and data available at http://www.itp.ucsb.edu.

\bibitem{kl}  G. Kotliar and J. Liu, Phys. Rev. B {\bf 38}, 5142 (1988).

\bibitem{dhl}  D.-H. Lee, Phys. Rev. Lett. {\bf 84}, 2694 (2000).

\bibitem{nayak}  C. Nayak, Phys. Rev. Lett. {\bf 85}, 178 (2000).

\bibitem{comment} A two-band model with a $d$-wave normal
state pseudogap has  also been considered by Benfatto {\it et al.},
Eur. Phys. J. {\bf 17}, 95 (2000).
There are two essential differences between this work
and Ref.\cite{clmn}. 1) Benfatto {\it et al} introduce
separate intra- and inter-band {\it d}-wave pairing. 2)
There is time reversal symmetry breaking in Ref.\cite{clmn}.

\bibitem{feynman}  See, for example, R. P. Feynman, {\it Statistical
Mechanics} (W. A. Benjamin, Inc. 1972) Chap. 10.
\end{references}
\end{document}